\documentclass[conference]{IEEEtran}

\usepackage{graphicx}
\usepackage{grffile}
\usepackage{amsmath}
\usepackage{booktabs}
\usepackage{stfloats}
\usepackage{url}
\usepackage[hidelinks]{hyperref}
\usepackage{array}
\usepackage[numbers,sort&compress]{natbib}
\usepackage[caption=false,font=footnotesize]{subfig}
\title{From Continuous Deployment to Queryable Dataset: Terabyte-Scale AIS-Aligned\\Passive Acoustic Labelling}

\author{
\IEEEauthorblockN{Wayne Renaud, Priyanka Aravindan, Gabriel Spadon}
    \IEEEauthorblockA{
        \textit{Faculty of Computer Science}, 
        \textit{Dalhousie University}, Halifax, Canada
    }\
    \{wayne.renaud, priyanka.aravindan, spadon\}@dal.ca
}

\newcommand{\fillpar}[1]{{\par\parfillskip=0pt\relax#1\par}}
\begin{document}
\maketitle

% Reduce vertical space around tables and figures
\setlength{\textfloatsep}{4pt plus 1pt minus 1pt}
\setlength{\floatsep}{4pt plus 1pt minus 1pt}
\setlength{\intextsep}{4pt plus 1pt minus 1pt}
\setlength{\dbltextfloatsep}{4pt plus 1pt minus 1pt}
\setlength{\abovedisplayskip}{4pt plus 1pt minus 1pt}
\setlength{\belowdisplayskip}{4pt plus 1pt minus 1pt}
\setlength{\abovedisplayshortskip}{2pt plus 1pt minus 1pt}
\setlength{\belowdisplayshortskip}{2pt plus 1pt minus 1pt}
\setlength{\abovecaptionskip}{3pt}
\setlength{\belowcaptionskip}{0pt}

\begin{abstract}
\parfillskip=0pt\relax Long-duration passive acoustic deployments produce large archives of recordings that are not linked to vessel tracks or encounter structure, leaving range and contact conditions unavailable as variables and requiring manual selection for analysis. To address this limitation, we propose a database-native workflow that aligns hydrophone recordings with Automatic Identification System (AIS) position reports to produce distance-resolved data. Fixed-duration recording windows and AIS messages are stored as persistent geospatial tables and associated through an indexed spatiotemporal join, replacing in-memory nested iteration with a single scalable set-based database process capable of handling continuous, multi-year, million-window archival deployments without exhausting available memory. In this study, the approach processes approximately $9.5\times10^5$ recording windows and $6.9\times10^6$ AIS position reports, producing a structured table that separates no-contact, single-contact, and two-contact windows, with the closest point of approach computed directly where applicable and background conditions characterized via deterministic spectral ranking. This formulation enables a GeoAI framework in which spatially indexed, queryable data become directly usable for machine learning. The resulting data product reveals predominantly noise-dominated conditions, with vessel contributions emerging mainly at shorter ranges, indicating that the task lies in extracting structure under background-limited regimes. Spectrogram and quantitative analyses show weak tonal signatures embedded in noise and a consistent decay of signal-to-noise ratio with distance, supporting the use of this representation for scalable machine learning, similarity analysis, and predictive acoustic modelling in real maritime environments.
\end{abstract}

\section{Introduction}

Passive acoustic monitoring now operates at scales that cannot be inspected manually. Learning-based methods continue to improve, but available datasets are either small or lack the contextual information required for controlled evaluation~\cite{irfanDeepShipUnderwaterAcoustic2021, santos-dominguezShipsEarUnderwaterVessel2016b}. Public datasets are typically curated for supervised classification, with limited preservation of source geometry and operational variability, restricting their use in studies that require explicit control over acoustic and spatial conditions.

When available, external tracking independently describes surface traffic. Aligning that data with fixed-duration recordings in memory does not scale. As a result, most automated passive acoustic monitoring systems focus on detecting vessel presence rather than constructing datasets tied to vessel geometry. For example, the DeepPAM~\cite{topplePassiveAcousticMonitoring2024} system detects and classifies ships from spectrogram inputs and compares predictions with AIS tracks as a proxy for ground truth. The output in such workflows is binary presence rather than a dataset with computable range and encounter structure.

\begin{figure}[htbp]
\centering
\includegraphics[width=\columnwidth]{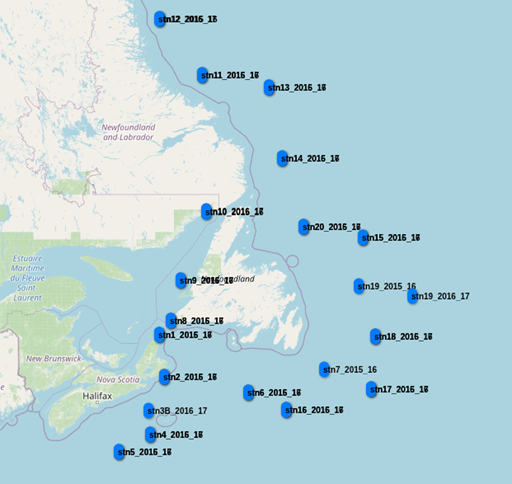}
\caption{Location of the fixed sensors used for the data construction. Each marker is a single station; the processing model is identical across stations.}
\label{fig:station_map}
\end{figure}

Most processing approaches rely on nested iterations over windows, vessels, and individual position reports. At archive scale, this results in tens of millions to trillions of comparisons, so full deployments cannot be completed in practice. For the deployment used in this study (see Figure~\ref{fig:station_map}), this corresponds to approximately $6.6\times10^{12}$ candidate window--AIS comparisons. To address this limitation, the database construction introduced here replaces hand-curated file selection and in-memory joining with a set-based, indexed spatiotemporal join. The experimental space is therefore defined by query rather than by manual inspection.

\looseness=-1 Accordingly, this paper introduces a database-based workflow that aligns passive acoustic recordings with AIS data to derive source range and contact condition as explicit, queryable variables. The construction allows subsets to be rebuilt from the archive under consistent definitions, enabling controlled, geometry-resolved evaluation beyond binary weak labelling. Here, AIS-conditioned labels act as contextual indicators rather than definitive acoustic ground truth, while training and test subsets can be formed at fixed range, fixed contact count, and ranked background condition.

The contribution of this work is a database-native construction that replaces nested iteration with a set-based spatiotemporal join executed directly within the database, enabling deployment-scale processing. The alignment with AIS data produces a distance-resolved weak labelling scheme in which single- and two-contact cases are defined from time-aligned vessel tracks and sensor position, preserving closest point of approach as a measurable quantity. An operational background class is defined from AIS-verified no-contact windows and structured through deterministic spectral ranking, providing a reproducible representation of the ambient acoustic field. Together, these components form a reproducible data product in which subsets can be constructed by query over range, contact count, and background condition.

\fillpar{The remainder of this paper is organized as follows. Section~\ref{sec:related_work} reviews existing underwater acoustic datasets and AIS-conditioned labelling approaches. Section~\ref{sec:methods} describes the database-native construction, contact classification, and background characterization workflow. Section~\ref{sec:results} presents the detailed physical and statistical characterization of the constructed single- and two-contact classes across the full deployment. Sections~\ref{sec:discussion}--\ref{sec:limitations} discuss the implications, reproducibility, and limitations of the workflow, and Section~\ref{sec:conclusion} concludes the paper.}

\section{Related Work}
\label{sec:related_work}

Existing underwater acoustic datasets have supported rapid progress in classifier design, but they were primarily constructed for supervised learning rather than deployment-scale analysis. Public corpora are typically event-curated, focusing on isolated examples rather than continuous records. DeepShip, for example, provides 47~h of labelled recordings from 265 vessels arranged as discrete events, each with a single dominant source inside a fixed radius~\cite{irfanDeepShipUnderwaterAcoustic2021}. ShipsEar follows the same event-based construction. Its database contains 90 recordings, each corresponding to a single labelled pass-by or background observation with a duration between 15~s and 10~min, resulting in an archive of isolated examples rather than a continuous deployment~\cite{santos-dominguezShipsEarUnderwaterVessel2016b}.

In ShipsEar, source--sensor separation is not derived from time-aligned vessel tracks and sensor position. Instead, the database stores only an approximate distance, reduced to four broad intervals ($<50$~m, $50$--$100$~m, $100$--$150$~m, $>150$~m) when exact values are unavailable. Consequently, range cannot be treated as a continuous variable or used to construct repeatable, fixed-range subsets for controlled propagation studies~\cite{santos-dominguezShipsEarUnderwaterVessel2016b}. In DeepShip, the labelling procedure assumes that only one vessel is present within the selection radius, excluding overlapping encounters during dataset construction~\cite{irfanDeepShipUnderwaterAcoustic2021}. Recordings are then divided into short segments, detaching the unit of analysis from the physical encounter and from its surrounding long-term acoustic context.

A similar pattern appears in more recent deep-learning frameworks for passive acoustic monitoring. The DeepPAM framework, for example, processes acoustic recordings from the Northern Watch arrays by segmenting time-series data and converting those segments into spectrogram inputs for convolutional neural network classification~\cite{topplePassiveAcousticMonitoring2024}. In these pipelines, acoustic segments serve as the training data rather than as a data product tied to vessel tracks and sensor geometry.

Several underwater acoustic classification studies rely on manually annotated datasets produced by expert operators, with labels derived through acoustic interpretation rather than a systematic, reproducible data construction pipeline~\cite{doanUnderwaterAcousticTarget2022}. Some datasets include background or ambient recordings as a separate class, reflecting that extended recording periods often contain no detectable vessel activity~\cite{doanUnderwaterAcousticTarget2022}.

\looseness=-1 Oceanship extends this work by ingesting Ocean Networks Canada (ONC) archives into a larger labelled corpus, reported as 121~h with 107,540 samples across 15 categories, with annotations derived from AIS fields and naming conventions~\cite{li2024oceanshiplargescaledatasetunderwater}. Yet the construction remains sample- or clip-based for target recognition rather than a regenerable dataset in which range and contact condition can be recomputed.

\fillpar{Collectively, these datasets support algorithm comparison, but they remove the continuous time context of long deployments, do not preserve source geometry as a queryable variable, and give no explicit representation of the underlying acoustic scene or its overlapping sources. Changing range definitions, contact conditions, or background selection therefore requires rebuilding the dataset rather than a query. Underwater acoustic classification thus continues to depend on small, often private collections, limiting open-science research~\cite{s22062181}.}

\looseness=-1 Recent surveys likewise note that the limited availability of annotated underwater acoustic datasets reduces reproducibility and slows progress~\cite{s22062181}. Many originate from operational or restricted contexts and cannot be independently reconstructed, and even when released, vessel identity fields are often removed or anonymized, limiting verification~\cite{li2024oceanshiplargescaledatasetunderwater}.

\looseness=-1 Tracking-conditioned weak labels exist, but most pipelines remain deployment-specific or limited to small subsets by computational cost. Machine-learning classifiers are thus often evaluated without explicit treatment of propagation, even though environmental factors directly affect the received signal and effective signal-to-noise ratio (SNR)~\cite{s22062181}. Robustness is frequently assessed using synthetic noise rather than measured background conditions~\cite{Capse}.

Table~\ref{tab:comparison} compares the capabilities of conventional underwater acoustic datasets with those of the data product produced by the proposed database-native construction. While existing datasets are distributed as fixed collections of labelled recordings, the proposed approach generates a persistent, queryable data product whose properties arise directly from the underlying construction methodology.

\begin{table}[t]
    \centering
    \caption{Comparison of conventional underwater acoustic datasets with the data product produced by the proposed database-native construction.}
    \label{tab:comparison}
    \setlength{\tabcolsep}{3pt}
    \begin{tabular}{@{}>{\raggedright\arraybackslash}p{2.4cm}
                    >{\raggedright\arraybackslash}p{2.7cm}
                    >{\raggedright\arraybackslash}p{2.7cm}@{}}
    \toprule
     & Curated datasets & Proposed data product \\
    \midrule
    Primary unit & Event clip & Fixed-duration window \\
    Archive retained & No & Yes \\
    Time continuity & Broken into selections & Full deployment \\
    Contact model & Single source & None / single-contact / two-contact \\
    Background class & Separate label or implicit & Operationally defined, queryable \\
    Range as variable & Metadata per clip & Column, queryable \\
    Scene complexity & Reduced at selection stage & Preserved in the data product \\
    Subset construction & Manual selection & Database query \\
    Experimental control & Limited & Range / contact / background stratified \\
    Regeneration from raw data & No & Yes \\
    \bottomrule
    \end{tabular}
\end{table}

\fillpar{Across these datasets, labels are attached to fixed collections of recordings or segments and cannot be recomputed under different definitions of range, contact condition, or background. In contrast, the construction presented here produces a persistent data product tied to the full sensor record, in which range, contact count, and background condition remain queryable variables. This shifts dataset construction from manual selection to a regenerable process defined over complete deployments.}

\begin{figure*}[b]
    \centering
    \includegraphics[width=.9\linewidth]{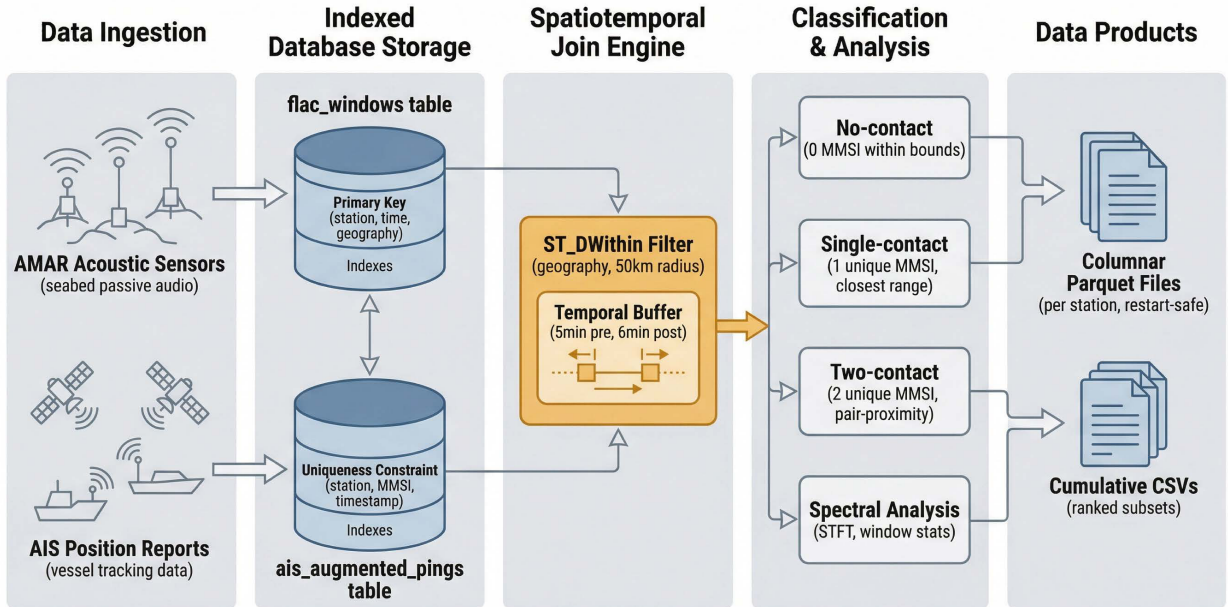}
    \caption{Methodology describing the database-native construction of the distance-resolved dataset. WAV recording windows and AIS position reports are stored as indexed tables and associated using a set-based spatiotemporal join. The result is a persistent table of contact windows with range-resolved context. \textit{The figure was designed by humans, partially generated by AI, but validated and edited by humans to ensure accuracy.}}
    \label{fig:wav_ais_join}
\end{figure*}

\section{Methods}
\label{sec:methods}

\looseness=-1 The methodology starts with acoustic recordings organized as fixed-duration windows within a common geodetic frame. These originate from seabed-deployed Autonomous Multichannel Acoustic Recorders (AMARs) manufactured by JASCO Applied Sciences~\cite{jasco_applied_sciences}. WAV metadata are stored in a persistent \textit{wav\_windows} table, while AIS reports are ingested and preprocessed using the AISdb framework~\cite{aisdb_docs} and stored in an \textit{ais\_augmented\_pings} table. Processing is performed once per station, after which the persistent tables are reused directly for all later analyses.

\begin{table}[b]
    \centering
    \caption{Distribution of contact classes for Station~17.}
    \label{tab:class_distribution}
    \begin{tabular}{lcc}
        \toprule
        Class & Count & Percentage \\
        \midrule
        No-contact & 24,727 & 69.30 \\
        Single-contact & 5,875 & 16.46 \\
        Two-contact & 5,077 & 14.23 \\
        \bottomrule
    \end{tabular}
\end{table}

All timestamps are converted to a common UTC-naive representation before processing. The end-to-end construction was validated using Station~17, the smallest station with a complete continuous record. Temporal joining uses a 5-minute pre-buffer, a 6-minute post-buffer, and a 50~km search radius. Contact classification is computed only within the recording window, while range is calculated using \texttt{ST\_Distance} following an initial \texttt{ST\_DWithin} filter. Outputs are written in columnar \textit{Parquet} format for downstream analysis. The overall workflow is illustrated in Fig.~\ref{fig:wav_ais_join}.

The joining is implemented as a set-based spatiotemporal query using \texttt{ST\_DWithin} on geography. A left join preserves windows without matching AIS reports, while post-processing verifies coordinate bounds and computed ranges.

Following the join, windows are classified into contact and background categories. A no-contact window contains no AIS-derived MMSIs within the specified spatiotemporal bounds and therefore represents the measured background acoustic state rather than acoustic silence.

\looseness=-1 Single-contact windows contain one unique MMSI within the recording interval, from which the closest point of approach (CPA), associated MMSI, and timestamp are extracted. Two-contact windows contain two unique MMSIs, with CPA computed per vessel and summarized by a pair-proximity metric. Contact subsets are aggregated by MMSI, and no-contact windows come directly from the join.

Background windows are characterized using full-window spectral analysis. Short-time Fourier transform statistics are averaged over each recording, normalized, filtered using percentile-based outlier rejection, and combined into a composite score used to rank background conditions.

\begin{figure*}[t]
    \centering

    \makebox[\textwidth][c]{%
        \subfloat[No-contact (high flow)]{%
        \includegraphics[width=0.31\textwidth]{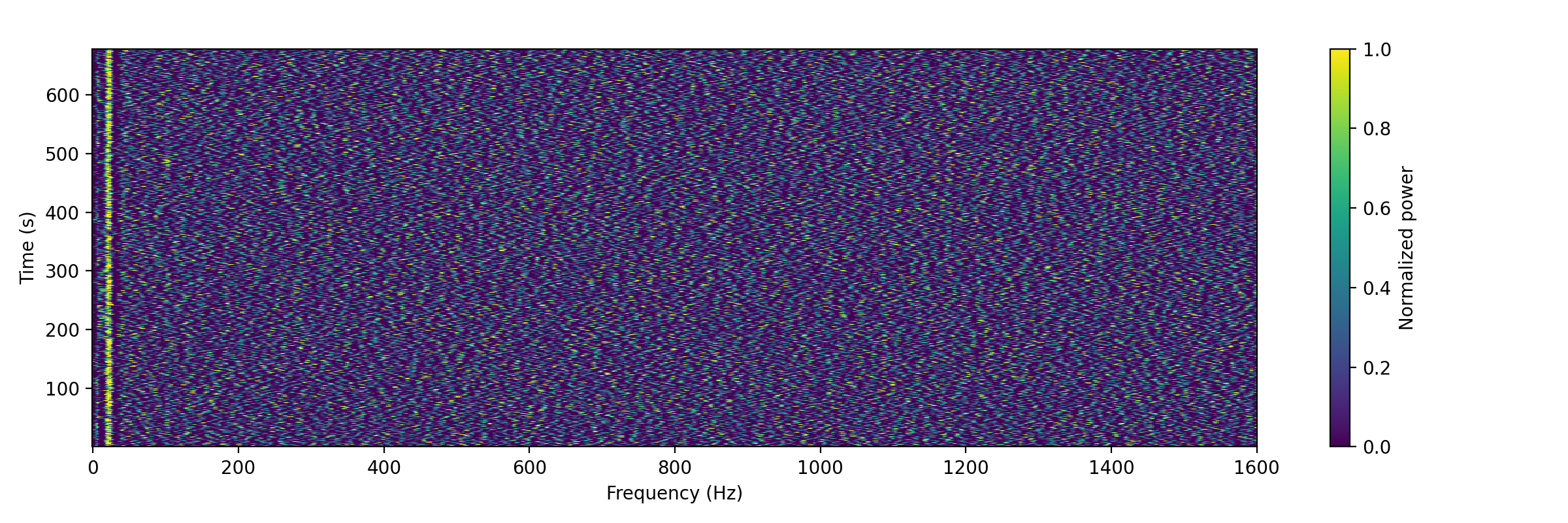}}%
        \hfill
        \subfloat[No-contact (low flow)]{%
        \includegraphics[width=0.31\textwidth]{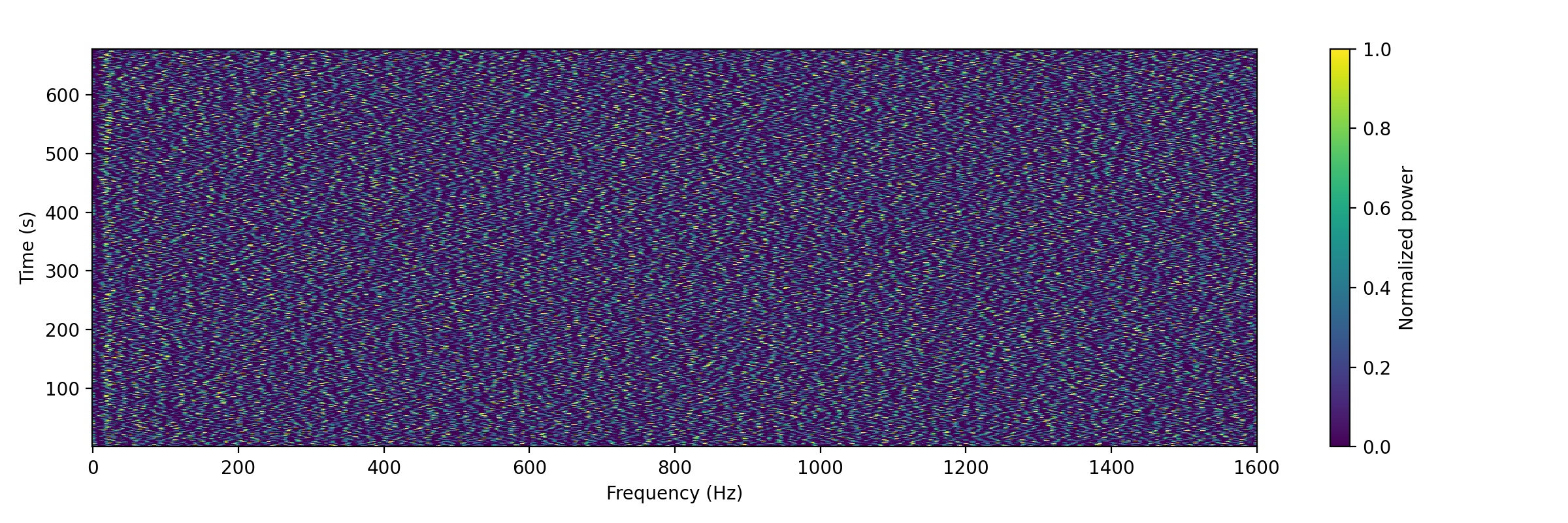}}%
        \hfill
        \subfloat[Single-contact (first CPA)]{%
        \includegraphics[width=0.31\textwidth]{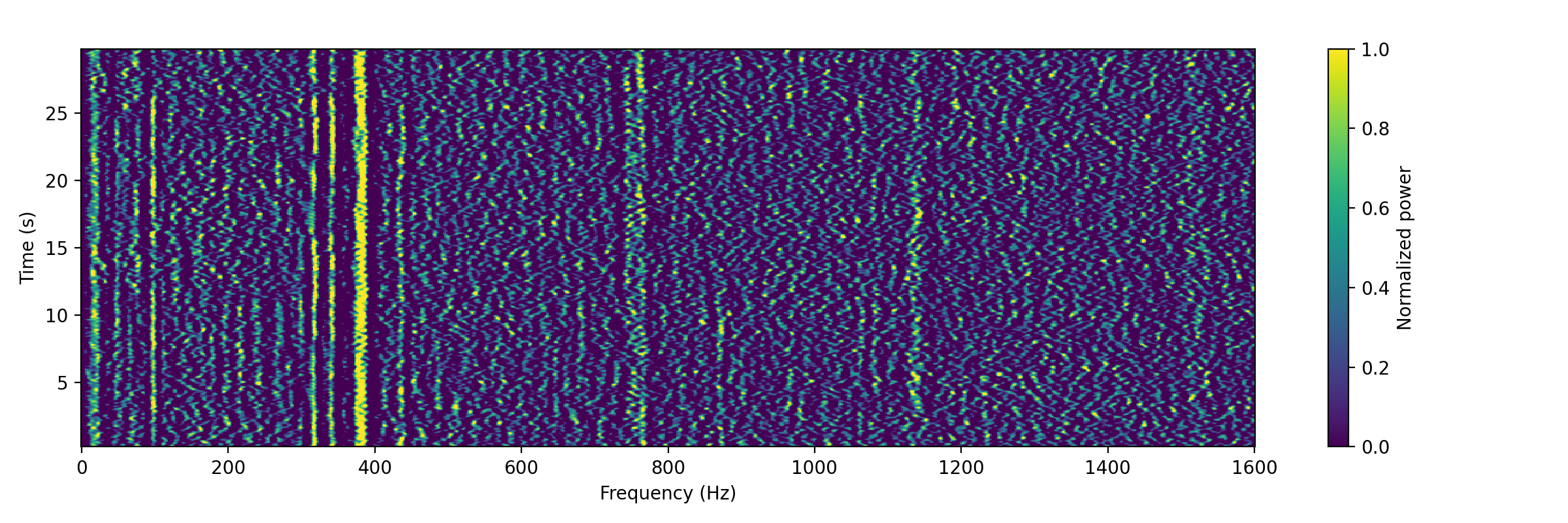}}%
    }

    \vspace{0.5em}

    \makebox[\textwidth][c]{%
        \subfloat[Single-contact (second CPA)]{%
        \includegraphics[width=0.31\textwidth]{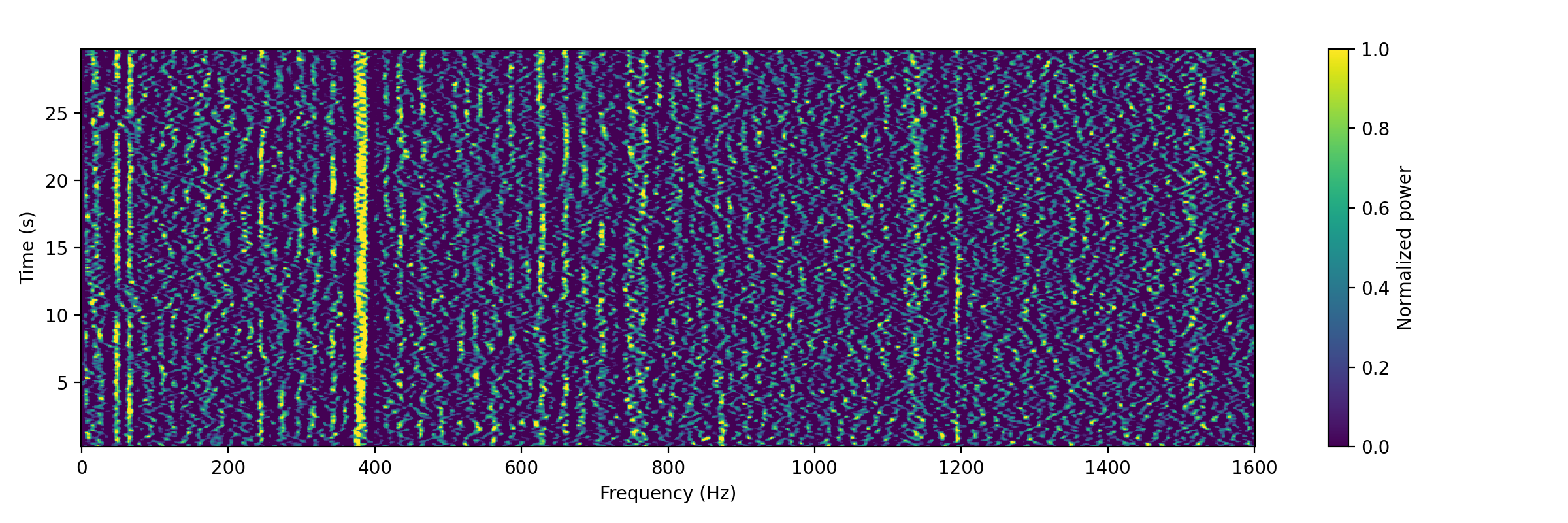}}%
        \hfill
        \subfloat[Two-contact (first vessel at CPA)]{%
        \includegraphics[width=0.31\textwidth]{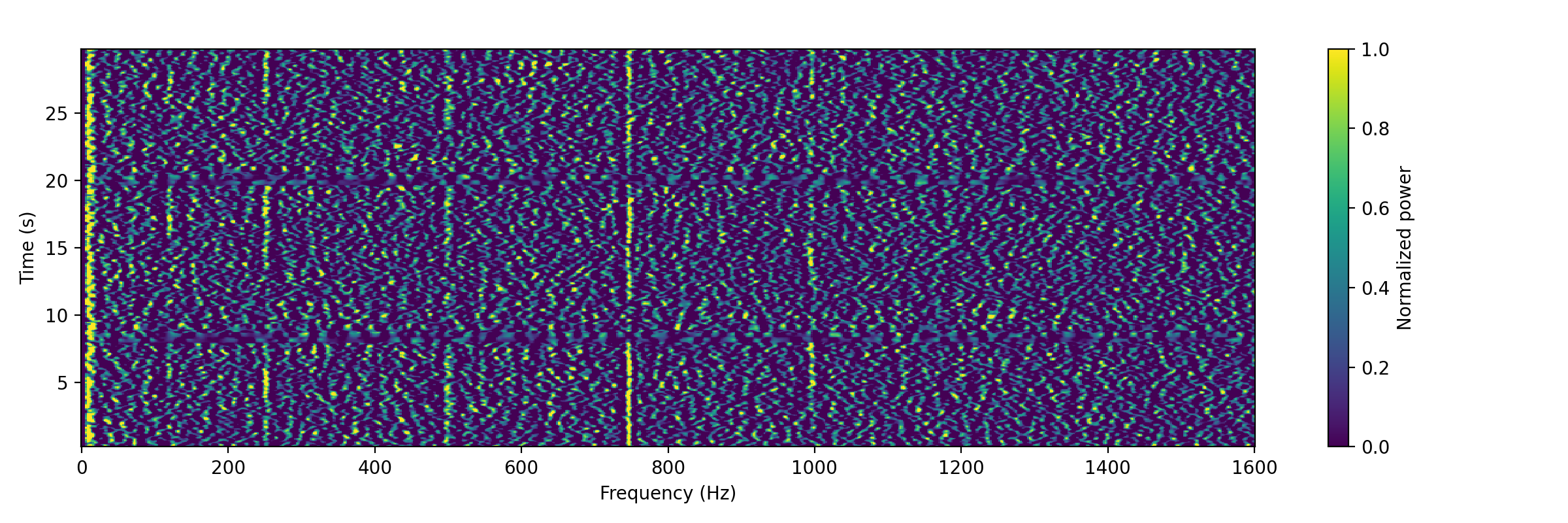}}%
        \hfill
        \subfloat[Two-contact (second vessel at CPA)]{%
        \includegraphics[width=0.31\textwidth]{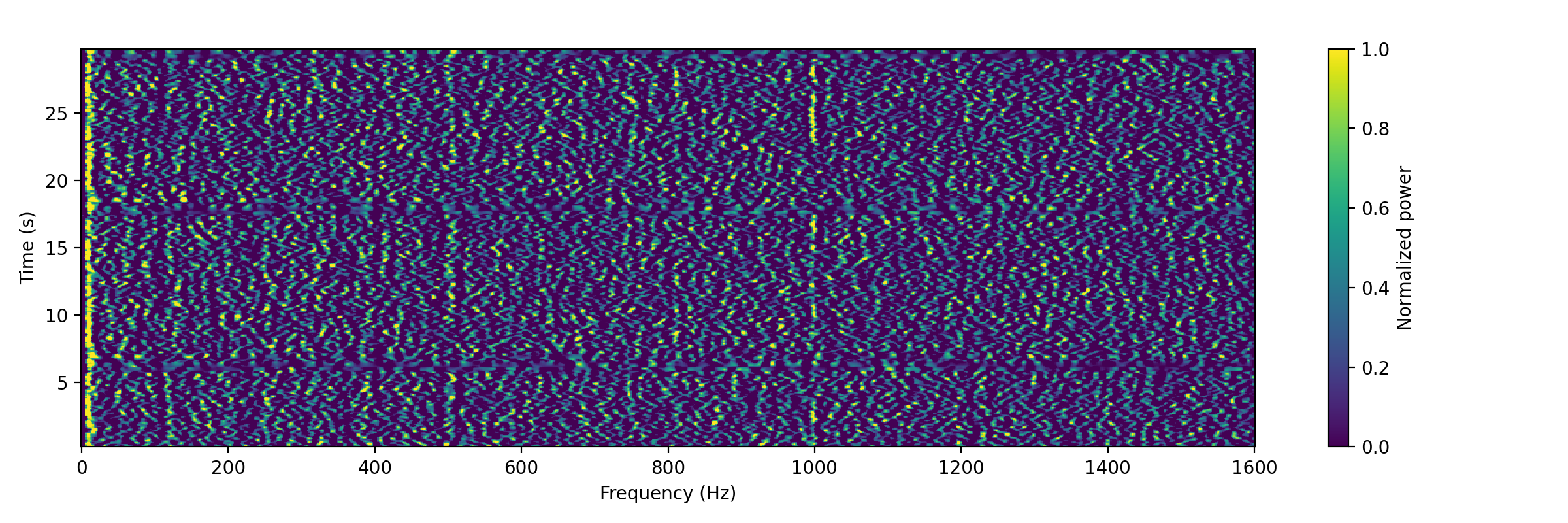}}%
    }

    \caption{
    Representative spectrograms across contact conditions. (a)--(b) no-contact windows showing variability in background and flow noise. (c)--(d) single-contact examples at different ranges and relative geometries with respect to the sensor, illustrating vessel-dependent tonal structure. (e)--(f) two-contact examples with similar closest point of approach (CPA) ranges but different encounter geometries relative to the sensor, showing overlapping spectral contributions.
    }
    \label{fig:spectrogram_examples}
\end{figure*}

The operational cost is dominated by AIS table construction and the station-level spatial join. Building \textit{ais\_augmented\_pings} processed approximately $6.9\times10^6$ position reports and required about 12.5~h, while the validation join for Station~17 required about 10~h. These one-time processing costs produce a persistent data product that supports subsequent analyses without repeating the fusion process. The resulting contact-class distribution is summarized in Table~\ref{tab:class_distribution}.

\section{Results}
\label{sec:results}
\subsection{Physical Characterization of Contact Classes}

\looseness=-1 Representative spectrograms for the three contact conditions are shown in Fig.~\ref{fig:spectrogram_examples}. The 11-minute no-contact windows exhibit background-dominated structure with no persistent tonal components, although low-frequency flow noise varies across conditions. In contrast, the 30-second (15-second windows either side of the CPA time) single-contact windows reveal narrowband tonal features associated with vessel machinery, with structure varying across vessels even at comparable ranges. The two-contact cases exhibit overlapping spectral contributions from multiple nearby vessels, further increasing the overall spectral complexity.

These examples show that vessel signatures are not cleanly separable from the background by visual inspection. Instead, the observed structure is consistently present and varies with both vessel and range, suggesting it can be captured through statistical characterization rather than visual interpretation.

\begin{figure}[ht]
    \centering
    \includegraphics[width=0.9\columnwidth]{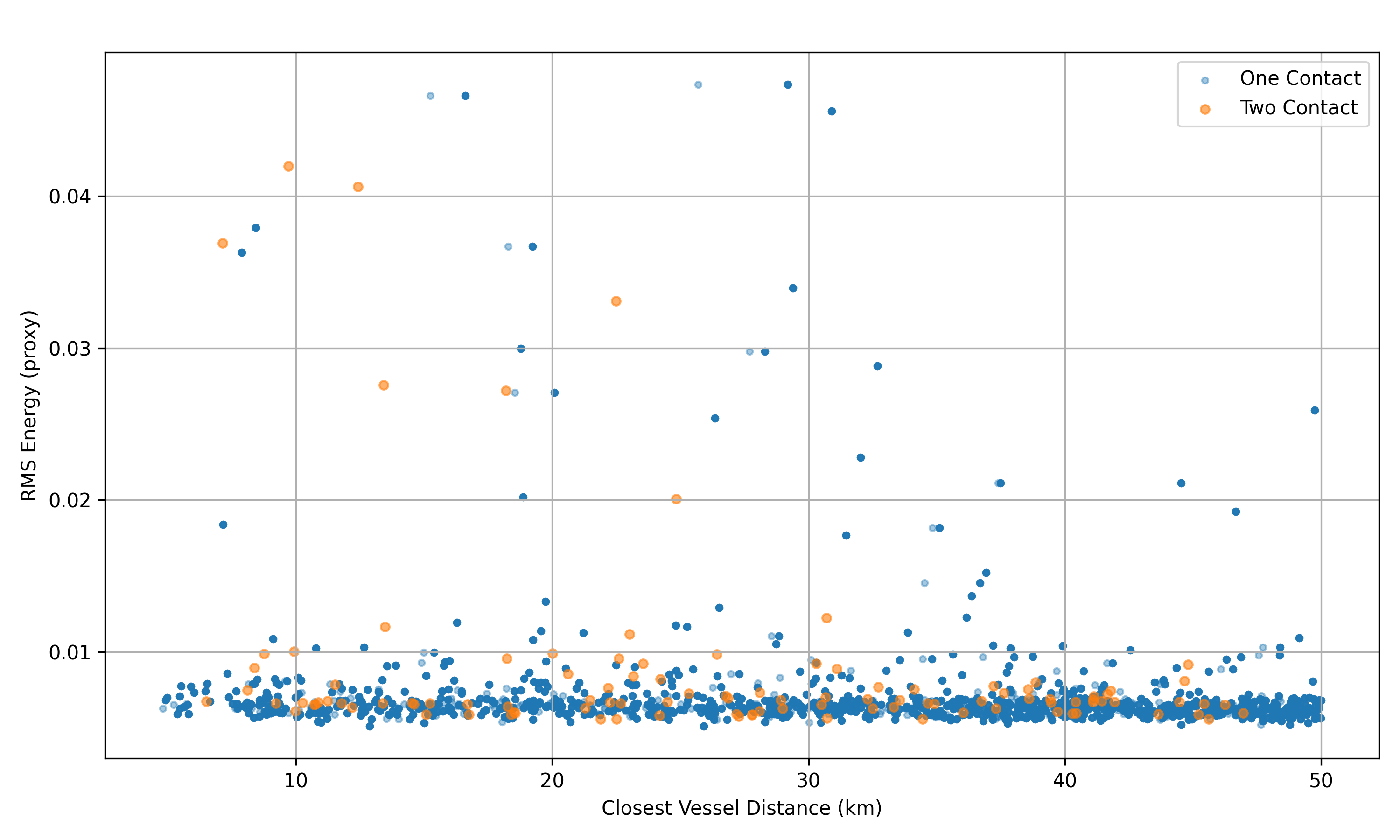}
    \caption{Root mean square (RMS) energy as a function of vessel distance for single- and two-contact windows. RMS increases at shorter ranges, consistent with propagation loss in underwater acoustic environments~\cite{urick, kinsler_fundamentals}, while most observations remain near the background noise level.}
    \label{fig:rms_distance}
\end{figure}

\fillpar{This relationship over full recording windows is shown in Fig.~\ref{fig:rms_distance}. Most observations lie within a narrow band near the background level, indicating that vessel contributions are embedded within the ambient acoustic field rather than clearly separated. RMS energy increases at shorter ranges, but varies considerably across individual vessels, reflecting systematic differences in vessel source level, hull design, propeller cavitation, loading, and instantaneous operating condition, so absolute levels are informative only in aggregate. The central challenge is therefore not detection under clean conditions, but extraction of structure in noise-dominated recordings where signal and background are weakly separated.}

% Remove the title from all images!!!

\begin{figure}[ht]
\centering
\includegraphics[width=0.9\columnwidth]{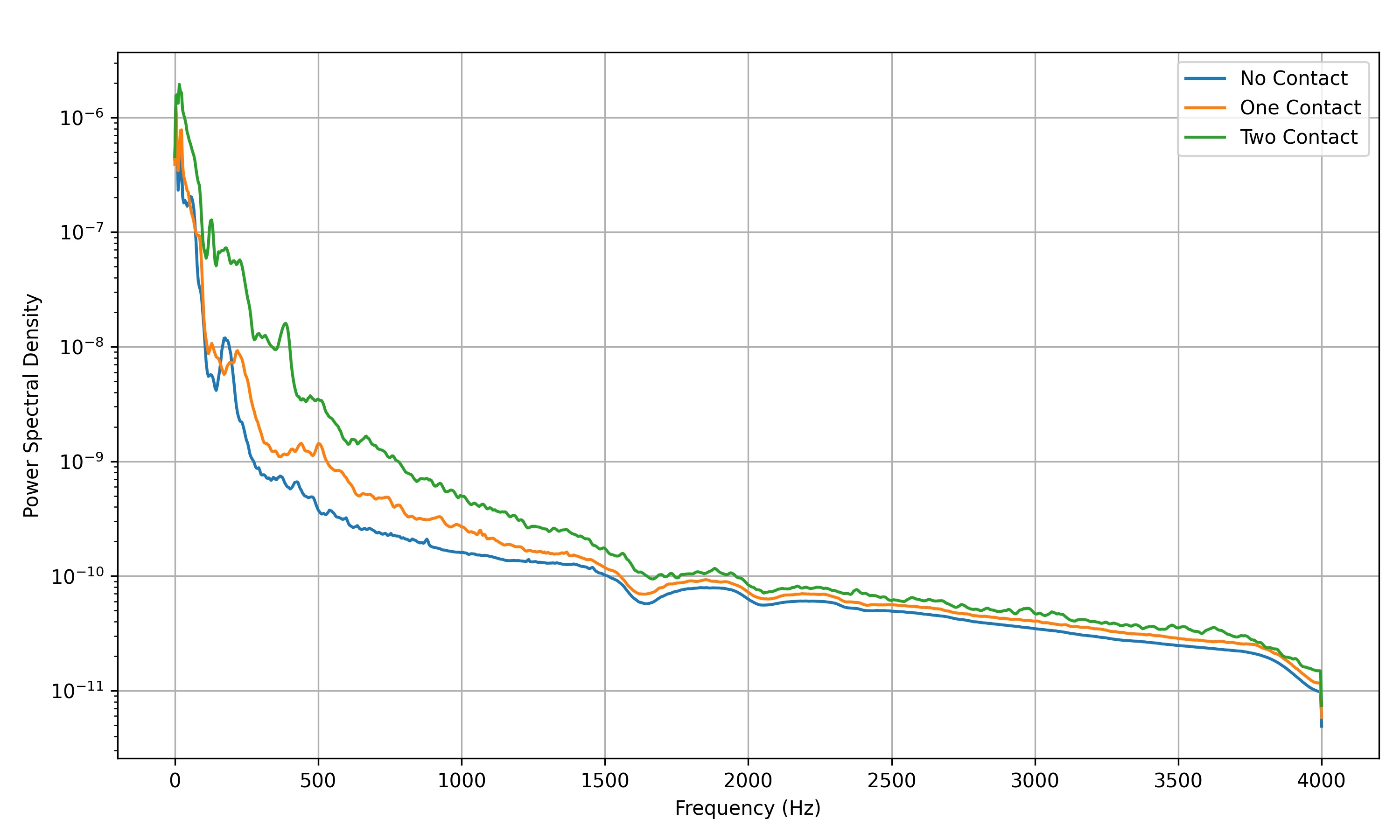}
\caption{Average power spectral density (PSD) for no-contact, single-contact, and two-contact windows. The separation is most evident at low frequencies, where vessel-generated energy increases with contact count, while higher frequencies remain dominated by background noise.}
\label{fig:psd_classes}
\end{figure}

\looseness=-1 The frequency-domain structure, represented by the average power spectral density, is shown in Fig.~\ref{fig:psd_classes}. Separation between contact classes is most pronounced at low frequencies, where vessel-generated energy contributes above the background, consistent with known characteristics of ship-generated noise and low-frequency propagation~\cite{urick}. At higher frequencies, the spectra converge, indicating dominance by background noise. Consistent with the RMS observations, the signal appears as a structured increase over the background rather than a clean separation, mainly at low frequencies. The distinction between single- and two-contact cases is statistical rather than absolute, reflecting overlapping contributions within a noise-dominated field.

\looseness=-1 The deployment spans two sensor configurations, with a recovery and redeployment occurring in July 2016. This transition introduces a change in signal scaling and instrument response, which likely contributes to the observed discontinuity in the RMS profile. The background structure therefore reflects both environmental variability and deployment-related effects.

\begin{figure}[ht]
    \centering
    \includegraphics[width=0.9\columnwidth]{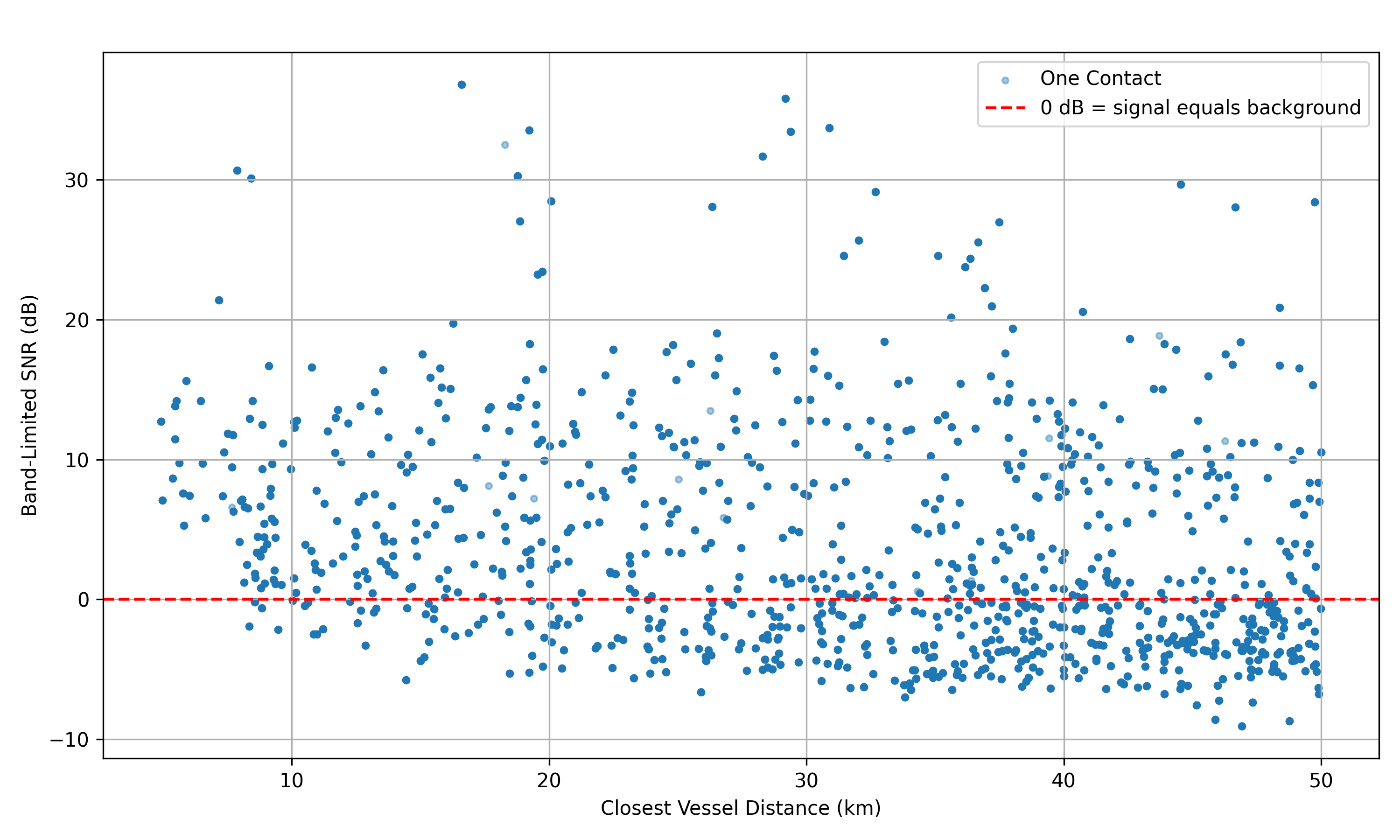}
    \caption{Band-limited signal-to-noise ratio (SNR) at CPA versus vessel range for single-contact windows. SNR is computed over 20--300 Hz and normalized by the median background level. Most observations are near the background, with a smaller subset of stronger sources showing higher contrast.}
    \label{fig:snr_scatter}
\end{figure}

\begin{figure}[ht]
    \centering
    \includegraphics[width=0.9\columnwidth]{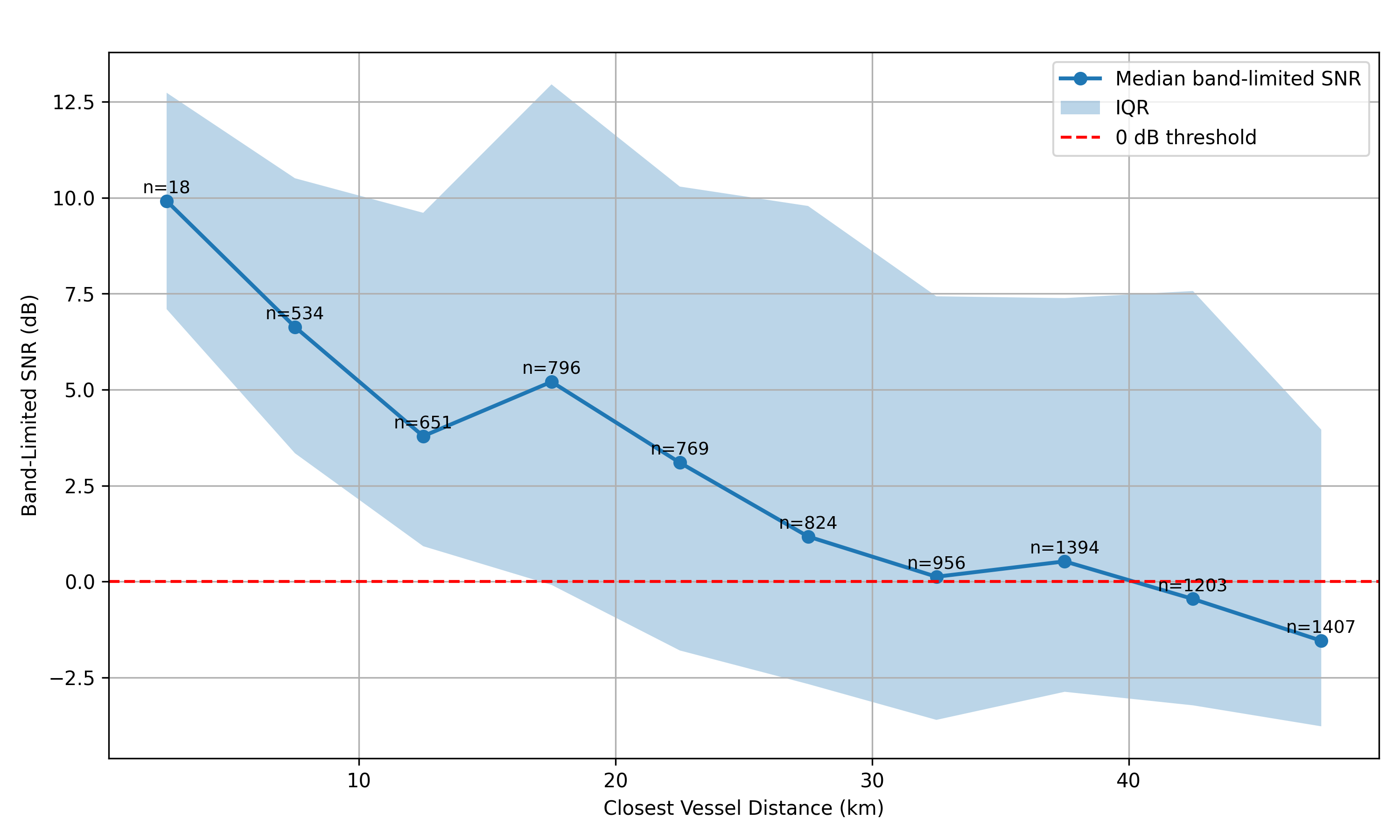}
    \caption{Median band-limited SNR versus range, with the interquartile range (IQR) shown as shading. The median approaches the background near 30--35 km, indicating a detection limit; the IQR reflects variability across the range.}
    \label{fig:snr_binned}
\end{figure}

\fillpar{The relationship between vessel range and acoustic detectability relative to the background for single-contact windows at closest point of approach (CPA) is shown in Fig.~\ref{fig:snr_scatter} and Fig.~\ref{fig:snr_binned}. Most observations lie within a narrow band near the background level, indicating that vessel signatures are typically weak relative to the ambient acoustic field even at CPA. A subset of vessels produces higher contrast, reflecting stronger source levels. When aggregated by distance, the median signal-to-noise ratio decreases with range and approaches the background level at approximately 30--35~km, consistent with propagation loss~\cite{urick}. The wide interquartile range indicates that detectability is jointly governed by range and source strength.}

These results indicate that the constructed dataset preserves the expected relationships between propagation, source variability, and background noise.

\subsection{Statistical Validation of Contact Classes}

\begin{figure}[ht]
\centering
\includegraphics[width=0.9\columnwidth]{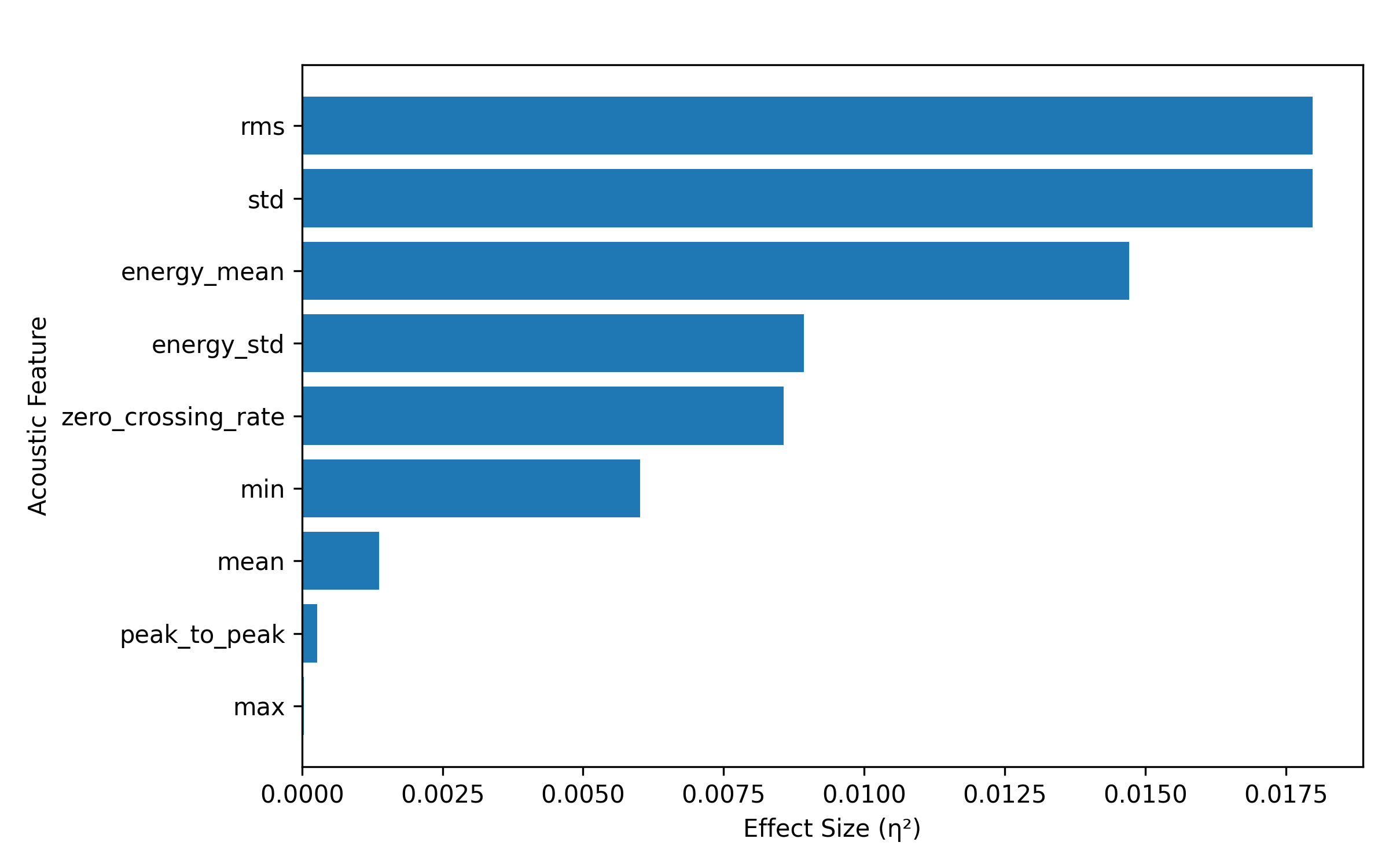}
\caption{Feature discriminative power across contact classes measured by ANOVA effect size ($\eta^2$). RMS, standard deviation, and energy-based features show the strongest separation, while others contribute less to class labelling.}
\label{fig:anova}
\end{figure}

The statistical separability of the contact classes is quantified in Fig.~\ref{fig:anova}. Discriminative power is concentrated in a small subset of features, most notably RMS, standard deviation, and energy-based measures, which capture the underlying signal energy and its temporal variability. This behaviour is consistent with vessel acoustics, where propulsion and machinery generate persistent tonal and broadband components that elevate energy over sustained intervals. In contrast, features such as maximum amplitude and peak-to-peak range exhibit limited separation because they are primarily influenced by transient excursions and do not reliably represent the continuous nature of vessel-induced acoustic signatures.

Although the distributions overlap at the sample level, their statistical structure remains consistently separable across features. Together with the ANOVA results, these observations indicate that the contact classes are not arbitrary labels but reflect measurable, reproducible differences in the acoustic data. The analysis therefore supports that the data construction preserves physically meaningful distinctions between no-contact, single-contact, and two-contact conditions within a noise-dominated acoustic field.

\section{Discussion}
\label{sec:discussion}

\fillpar{The primary contribution is a persistent, queryable dataset that supports physically defined subsets without repeating the costly fusion stage.}

The background-only subset is defined as windows without AIS-correlated surface traffic, without assuming the absence of other contributors. Untracked vessels, biological sources, and weather-driven noise may still be present, so the aim is to characterize variability rather than remove it. Statistical screening yields a reproducible representation of the ambient field, supporting controlled contact-in-noise experiments, synthetic injection, calibration, and stress testing. Because the environment is defined by measurement rather than modelling, comparisons hold across sensors despite propagation effects such as attenuation, multipath, and masking~\cite{urick, kinsler_fundamentals}.

\fillpar{More broadly, the construction shifts away from curated data collection. Rather than selecting isolated examples under known conditions, the dataset comes directly from operational recordings, preserving the variability, interference, and noise of real deployments while giving an explicit representation of scene complexity.}

This distinction affects the development and evaluation of learning-based methods. Whereas most approaches treat background as an uncontrolled nuisance on curated datasets, our data product retains the full deployment structure and exposes range, contact condition, and background state as queryable variables. Training and test sets can thus be built under matched conditions, including fixed range intervals, consistent background, or specific contact configurations, for example training at short range and testing where the signal-to-background ratio approaches unity. The result is not a single benchmark but an experimental substrate for repeatable, physically grounded evaluation across many learning tasks.

\section{Reproducibility}
\label{sec:reproducibility}

\fillpar{The workflow is deterministic under identical inputs and parameters. Persistent SQL tables are built once and reused across runs, while the spatiotemporal join is defined by fixed temporal and spatial constraints. Outputs are written in columnar \textit{Parquet} format, enabling selective reprocessing of individual stations or time intervals without rebuilding the entire data product.}

\section{Limitations}
\label{sec:limitations}
\fillpar{This work focuses on data construction and does not extend to acoustic ground-truth validation. The labels derive from external tracking rather than direct acoustic attribution, and should be read as contextual indicators rather than definitive ground truth. Their accuracy depends on the completeness of AIS reporting and on interpolation between position updates. Vessels without AIS transmission, including small craft and non-compliant platforms, are therefore not represented in the contact classes.}

A no-contact window denotes the absence of AIS-correlated surface vessels within the defined bounds, not acoustic silence, since untracked vessels, biological activity, and environmental noise may still be present. This uncertainty grows in multi-contact conditions, where overlapping vessels and propagation effects decouple geometric position from acoustic energy.

The labels are therefore weak and may carry temporal and spatial uncertainty. The work does not assess downstream detection, classification, or separation; its contribution is a reproducible, queryable dataset that preserves physically meaningful structure and enables such analyses.

\section{Conclusion}
\label{sec:conclusion}
We presented a database-native, archive-scale workflow for AIS-conditioned labelling of passive acoustic data. The distance-resolved dataset preserves expected relationships between propagation, source variability, and background noise while making range, contact condition, and background state explicitly and directly queryable.

The product retains measured variability, interference, and propagation effects, with range, contact count, and background state as explicit coordinates, providing a foundation for learning-based methods at deployment scale where models are trained and evaluated under physically defined regimes for systematic detection, separation, and representation learning.

\bibliographystyle{IEEEtran}
\bibliography{refs.bib}

\end{document}